\documentclass[conference, letter]{IEEEtran}
\IEEEoverridecommandlockouts

\ifCLASSINFOpdf
  \usepackage[pdftex]{graphicx}
  \DeclareGraphicsExtensions{.pdf,.jpeg,.png}
\else
\fi

\usepackage{cite}
\usepackage{amsmath,amssymb,amsfonts}
\usepackage{algorithmic}
\usepackage{graphicx}
\usepackage{textcomp}
\usepackage[left=1.29cm, right=1.29cm, top=1.9cm, bottom=4.29cm]{geometry}
\usepackage{enumitem}
\usepackage{anyfontsize}
\usepackage[font=small]{caption}
\usepackage{cite}
\usepackage{fancyhdr}
\usepackage{siunitx}

\makeatletter
\renewcommand{\fnum@figure}{Figure \thefigure}
\makeatother

\title{Impact of Traffic-Following on\\
Order of Autonomous Airspace Operations}

\author{\IEEEauthorblockN{Anahita Jain}
\IEEEauthorblockA{\textit{Metis FRA at NASA Ames Research Center} \\
\textit{The University of Texas at Austin}\\
Austin, USA\\
anaj18@utexas.edu}
\and
\IEEEauthorblockN{ Husni R. Idris}
\IEEEauthorblockA{
\textit{NASA Ames Research Center}\\
Moffett Field, USA\\
husni.r.idris@nasa.gov }
\and
\IEEEauthorblockN{ John-Paul Clarke}
\IEEEauthorblockA{\textit{The University of Texas at Austin} \\
Austin, USA\\
johnpaul@utexas.edu}
}

\renewcommand\IEEEkeywordsname{Keywords}
\IEEEaftertitletext{\vspace{-1\baselineskip}}

\begin{document}

\maketitle
\thispagestyle{fancy}

\noindent \begin{abstract} 

In this paper, we investigate the dynamic emergence of traffic order in a distributed multi-agent system, aiming to minimize inefficiencies that stem from unnecessary structural impositions. We introduce a methodology for developing a dynamically updating traffic pattern map of the airspace by leveraging information about the consistency and frequency of flow directions used by current as well as preceding traffic. Informed by this map, an agent can discern the degree to which it is advantageous to follow traffic by trading off utilities such as time and order. We show that for the traffic levels studied, for low degrees of traffic-following behavior, there is minimal penalty in terms of aircraft travel times while improving the overall orderliness of the airspace. On the other hand, heightened traffic-following behavior may result in increased aircraft travel times, while marginally reducing the overall entropy of the airspace. Ultimately, the methods and metrics presented in this paper can be used to optimally and dynamically adjust an agent's traffic-following behavior based on these trade-offs.

\end{abstract}

\vspace{0.3cm}


\section{Introduction}

In order to accommodate growing demand and improve safety, the airspace system is anticipated to increasingly include autonomous vehicles that interact collectively and integrate with other traffic sharing the same airspace. Initially, applications will include non-passenger scenarios, such as fire fighting or cargo delivery, using uncrewed aerial vehicles of different sizes. Eventually, the scope will expand to passenger-carrying vehicles for urban or regional air mobility. Successful implementation of these collective autonomous multi-vehicle systems, requires us to \textit{identify} key characteristics that make these systems ``good" in terms such as safety and efficiency. After that, identifying metrics to \textit{quantify} these characteristics becomes crucial, followed by the design of agent behaviors and coordination mechanisms to ensure we \textit{achieve} these desired traits in a collective autonomous multi-agent system.

To identify characteristics in the context of air mobility, we interviewed pilots to gain insights on good collective autonomous behavior, particularly under distributed visual flight rules. A key takeaway from these discussions is that pilots, with a primary focus on risk mitigation, actively seek to minimize uncertainty while integrating with other traffic. Examples of such efforts include planning to fly along commonly filed routes and flown procedures, following aircraft flown by more experienced operators, and following aircraft when finding paths through weather systems. Pilots also integrate into airport arrival flows while negotiating their sequence on the common communication frequency and merging into the traffic pattern. 

In air traffic management, fostering safe, orderly, and expeditious traffic are key objectives \cite{b1}. Air traffic controllers ensure safe separation distances between vehicles and often expedite traffic to enhance system throughput. In maintaining the orderliness of traffic, controllers ensure aircraft compliance with established route structures and procedures, and they apply equitable first-come, first-serve service. They also dynamically organize traffic into patterns to effectively navigate complexity and workload challenges, especially in airspaces with high traffic densities \cite{b18}\cite{b17}. Order is often traded for expediency and safety, for example, by deviating from the route structure to gain time or resolve traffic conflicts.

Considerable efforts have focused on ensuring safety in distributed autonomous systems, through automated collision avoidance systems, detect-and-avoid technology \cite{btcas}, and strategic self-separation concepts \cite{baop}. There has also been ample research in collaborative and distributed traffic flow management concepts and technologies to ensure user priorities are traded effectively with the collective safety and throughput of the airspace \cite{btasar} \cite{bctfm}. In this paper we focus on the characteristics of orderly traffic, a topic that has received relatively less attention but which we believe will be increasingly critical for scaling collective multi-vehicle autonomous systems to higher densities. In addressing the need for orderly traffic behavior, the existing body of research has delved into airspace complexity metrics from an air traffic controller workload perspective such as in \cite{pk-dyndensity} or, more intrinsically, such as in \cite{delahaye-entropy}. There were efforts to study the emergence of orderly traffic patterns in distributed settings using flexibility metrics designed to mitigate high risk \cite{idris-delahaye} \cite{idrisaviation2011}. It was shown that airspace complexity is reduced under such distributed schemes. Attempts were made to mimic collective behavior from other social species such as ants for traffic management \cite{durant}. Convoy formation has also been studied in the context of maintaining vehicle-to-vehicle coherence and alignment for energy saving and benefits such as flow structuring \cite{ishihara} \cite{Xu}. 



\begin{figure}[hbt!]
\centering
\includegraphics[width=0.3\textwidth]{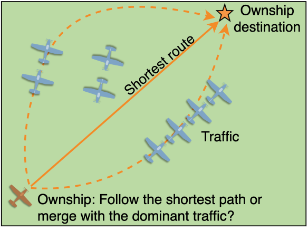}
\caption{An ownship can decide the degree to which it follows traffic, and which traffic it follows, as it progresses towards its destination.}
\label{following}
\end{figure}

In this paper, we investigate the dynamic emergence of traffic structure in a distributed multi-agent system. In doing so, we attempt to minimize the inefficiencies stemming from applying structure unnecessarily \textit{a priori}. We introduce a methodology for developing a traffic pattern map of the airspace by leveraging information about the consistency and frequency of flow directions used by current as well as preceding traffic. Informed by this map, an agent can discern the degree to which it is advantageous to follow traffic by trading off utilities such as time and order, as illustrated in Fig. \ref{following}. When agents follow traffic, the outcome is the dynamic emergence or formation of specific paths, or ``airways", that are used more frequently than others without pre-imposed route structures.

This technique also enables an agent to customize its degree of traffic-following behavior based on trading its own priorities with the overall traffic concerns. For example, while traffic-following behavior might come at an expense to aircraft, for instance, in terms of travel time, the increased order afforded by this strategy ultimately contributes to a more manageable and safe airspace environment. In this paper, we quantify the trade-off between order and time under low-density conditions, which provides a base for quantifying the impacts of order on airspace operations.


The paper is organized as follows: Section \ref{sec:Method} covers the modeling framework, path planning algorithm, and quantitative metrics for measuring order. Section \ref{sec:results} presents simulation results demonstrating the emergent traffic order and its effect on travel time, and Section \ref{sec:future} concludes this paper outlining avenues for further research.

\section{Methodology}\label{sec:Method}

In order to study collective autonomous behavior that aims to maintain orderly traffic in a distributed manner, we developed the following models, algorithms and metrics:
1. A map of the airspace that depicts the traffic pattern based on information from preceding traffic in Section \ref{sec:map}. We assume that this information is available to the vehicles either through their own sensors covering the airspace region of interest or through a service that gathers the information and broadcasts it to all vehicles. 
2. A cost function for each agent to use the traffic pattern map information and calculate the amount of traffic-following behavior to apply relative to other utilities in Section \ref{subsec:concepts}. 
3. A path planning algorithm that minimizes the cost function in Section \ref{subsec:strategy}.
4. A metric to measure the order of the traffic based on entropy in the traffic direction in Section \ref{entropy}.

\vspace{-0.1cm}
\subsection{Traffic pattern map}\label{sec:map}
\begin{figure}[hbt!]
\centering
\includegraphics[width=0.3\textwidth]{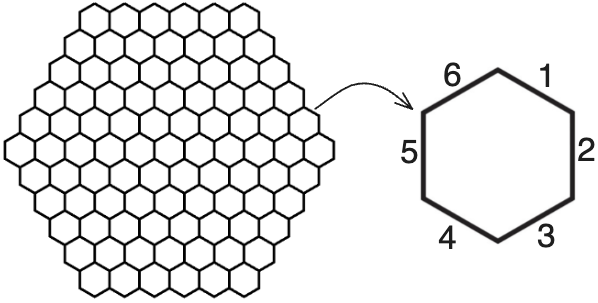}
\caption{
We partition the airspace into a hexagonal grid, where each cell is assigned edges numbered from 1 to 6. This numbering system facilitates the indexing of the cost matrix associated with each cell.}
\label{hex}
\end{figure}
We assume the airspace is two-dimensional and partition it into regular hexagons, tiled and pairwise congruent \cite{patel}, as depicted in Fig. \ref{hex}. Since our objective is to capture traffic directions, a hexagonal grid allows us to track more directions than, for example, a square grid, at a reasonable computation cost. The edges of each hexagon are numbered as shown in the figure. For each possible traversal of a flight through the hexagon, we specify a corresponding ordered pair of edges, in the format (entry edge, exit edge). Such a pair will be referred to as an \textit{edge pair} for the given flight. Thus, for each hexagon, a total of 36 edge pairs are possible. The number of traversals through an edge pair $(i, j)$ within a hexagon will be denoted by $t_{i, j}$. Thus, for each hexagon, a 6$\times$6 matrix denoted by Eq. \ref{t_matrix} records the number of aircraft that have traversed this edge pair so far, where the diagonal entries represent U-turns. 
\begin{equation}
\label{t_matrix}
T = 
\begin{bmatrix}
    t_{1,1} & t_{1,2} & t_{1,3}& t_{1,4}& t_{1,5}& t_{1,6}\\
    t_{2,1} & t_{2,2} & t_{2,3}& t_{2,4}& t_{2,5}& t_{2,6}\\
    t_{3,1} & t_{3,2} & t_{3,3}& t_{3,4}& t_{3,5}& t_{3,6}\\
    t_{4,1} & t_{4,2} & t_{4,3}& t_{4,4}& t_{4,5}& t_{4,6}\\
    t_{5,1} & t_{5,2} & t_{5,3}& t_{5,4}& t_{5,5}& t_{5,6}\\
    t_{6,1} & t_{6,2} & t_{6,3}& t_{6,4}& t_{6,5}& t_{6,6}\\
\end{bmatrix}
\end{equation}
For example, if a cell has had no aircraft traversals in the past, the traffic matrix, T, is simply a 6$\times$6 zero matrix. Then, by time $\Delta $ if an aircraft has entered via the second edge and exited via the fifth and another aircraft has entered via the fourth and exited via the first, the updated corresponding T matrix of the cell is: 
\begin{center}
$ T = \begin{bmatrix}
    0 & 0 & 0 & 0 & 0 & 0\\
    0 & 0 & 0 & 0 & 1 & 0\\
    0 & 0 & 0 & 0 & 0 & 0\\
    1 & 0 & 0 & 0 & 0 & 0\\
    0 & 0 & 0 & 0 & 0 & 0\\
    0 & 0 & 0 & 0 & 0 & 0\\
    \end{bmatrix}
$
\end{center}
\vspace{0.05cm}

It is important to note here that non-zero entries signify the passage of traffic through the corresponding entry-exit pairing, with the value indicating the number of aircraft that have used the same pathway. By using this technique, we capture not only the different directions of traffic through a cell, but also the number of aircraft that are traversing the hexagon in these directions. Additionally, in the current setup, traffic entries are not differentiated on the basis of the when they occurred. In future extensions, a time-decay factor may be introduced to discount older information relative to newer information. 
\subsection{Cost function model} \label{subsec:concepts}

In the distributed environment in this paper, any aircraft, called ``ownship", makes its own path-planning decision using a cost function. An ownship incurs a cost whenever it passes through a hexagon from any one edge to another. This cost arises from fuel burn, winds, and weather conditions, independent of any traffic considerations. This will be referred to as the \textit{unimpeded transit cost}. Additionally, if other flights are traversing the grid, the necessity to share the airspace results in an additional cost to the ownship. This will be referred to as the \textit{traffic cost}. The sum of these two costs will be called the \textit{total cost of transit} from any one edge of a cell to another. These are detailed in the following subsections.

\subsubsection*{Unimpeded transit cost through a cell}
The cost of unimpeded transit through an edge pair  $(i, j)$ is denoted by $u_{i, j}$. It can be adjusted based on how expensive it is for an ownship to transit an edge pair based on winds, weather, and fuel burn within the airspace. For example, edge pairs encompassing bad weather are set up to have higher prices of transit to deter aircraft from traveling through there. Therefore, less favorable transit pairs are set up to have higher costs and vice-versa. For all simulations conducted for this paper, the unimpeded cost of transit through an edge pair is the simply distance between their midpoints, without any adjustments for wind and weather. In the case of U-turns, the cost is approximated to twice the size of a hexagon. 

For a hexagon, there are 36 unique values denoting transit cost between each edge pair. Again, for each individual hexagon, these are best stored in a 6 $\times$ 6 matrix as follows. 

\begin{equation}
    \label{u_matrix}
    U = 
\begin{bmatrix}
    u_{1,1} & u_{1,2} & u_{1,3}& u_{1,4}& u_{1,5}& u_{1,6}\\
    u_{2,1} & u_{2,2} & u_{2,3}& u_{2,4}& u_{2,5}& u_{2,6}\\
    u_{3,1} & u_{3,2} & u_{3,3}& u_{3,4}& u_{3,5}& u_{3,6}\\
    u_{4,1} & u_{4,2} & u_{4,3}& u_{4,4}& u_{4,5}& u_{4,6}\\
    u_{5,1} & u_{5,2} & u_{5,3}& u_{5,4}& u_{5,5}& u_{5,6}\\
    u_{6,1} & u_{6,2} & u_{6,3}& u_{6,4}& u_{6,5}& u_{6,6}\\
\end{bmatrix}
\end{equation}
\subsubsection*{Traffic cost through a cell}

There are a number of ways that traffic can affect an ownship. In this cost function we model the inclination of an aircraft to follow traffic patterns if beneficial to it. In the next subsection, we model how an aircraft can maintain minimum horizontal separation distances from other aircraft using a conflict resolution technique. 

For each edge pair, we define a traversal cost based on the number of aircraft that have used the pair. This cost is formulated as $(1 - k_t*{\hat{t_{i, j}}})$ where $k_t$ is the \textit{traffic-following factor} and $\hat{t_{i, j}}$ is the normalized entry from Eq. \ref{t_matrix}. Subtracting the normalized traffic count from 1 makes the edge pairs with more traffic in them less costly to the ownship, thus making pairs with higher traffic "attractive" to the ownship. The traffic-following factor, $k_t$ is adopted as a gain in order to tailor the degree to which traffic is attractive to an ownship. The higher the value of $k_t$, the less costly it becomes to use the edge pair. Therefore, the higher the value of $k_t$, the more inclined an ownship is to follow traffic. 

Similar to previous setups, for each hexagon, there are 36 such values, which are represented in a 6$\times$6 matrix. Hence, the traffic cost through different entry-exit pairs in a hexagon is given by: 


\begin{equation}
    \label{t_cost}
    \left[
\textbf{1}_{6x6} - k_t\frac{T}{\sum_{i, j}t_{i, j}}
\right]
\end{equation}

where $\textbf{1}_{6x6}$ is a 6 $\times$ 6 matrix of ones and $\sum_{i, j}t_{i, j}$ is the grand sum of the traffic matrix of that hexagon. 

The current approach for quantifying traffic costs considers the direction of aircraft and vehicular count. In future work, we plan to incorporate additional characteristics such as similarity between traffic by considering speeds and sizes of different agents. 

\subsubsection*{Total cost of transit through a cell} 
Summing up the unimpeded transit cost and the traffic costs of a cell, we get the total cost of transit through a cell as follows:

\begin{equation}
    \label{final_cost_equation}
    C = U
 +  \left[
\textbf{1}_{6x6} - k_t\frac{T}{\sum_{i, j}t_{i, j}}
\right]
\end{equation}

\subsection{Path planning and conflict resolution} \label{subsec:strategy}

\begin{figure}[hbt!]
\centering
\includegraphics[width=0.45\textwidth]{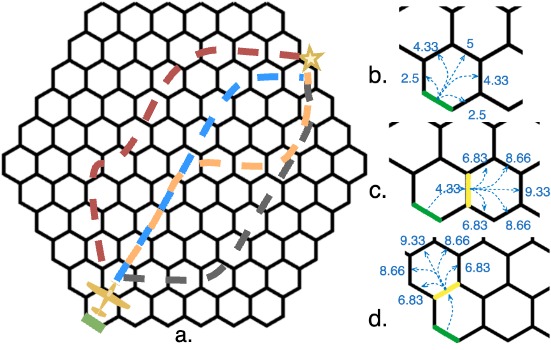}
\caption{(a). Illustrating different trajectories that a vehicle can take to get from one edge in the grid to another. (b). Using Dijkstra's algorithm, an aircraft computes the total cost for each possible path, starting from the initial edge. (c). The least costly edge that can lead to another hexagon is picked for the next cost step calculation. (d). Again, the least costly edge that has not been explored, starting from the base edge, is picked for the next cost step calculation. This continues until the next edge to be explored is the destination edge.} 
\label{severalpaths}
\end{figure}
There are generally multiple paths an ownship can take to get from one hexagon to another within the grid as illustrated in Fig.\ref{severalpaths}a. In this section we introduce an algorithm that uses the cost function defined in Section \ref{subsec:concepts} to plan an ownship's trajectory and a method for conflict resolution.

\subsubsection*{Least costly path calculation}
For every possible path between the initial and final position, we define a cost using the hexagonal costs introduced in Eq. \ref{final_cost_equation}.


We cast the path planning problem by introducing an undirected graph \cite{b16} whose nodes are all the edges of the hexagons\footnote{In the graph theoretic approach used here, the terms nodes and arcs are used, instead of the more commonly used terms vertices and edges, to avoid confusion with the hexagonal grid's geometry.}. There are only two sets of possible arcs defined in this setup. 1. Two nodes corresponding to two edges of the same hexagon are connected by an arc that equals the total cost of traversal of that edge pair. 2. Two nodes corresponding to the coincident edges of neighboring hexagons are connected by an arc that equals the cost of 0. 

An ownship computes its trajectory using this weighted graph setup. To calculate the least costly path, we now use Dijkstra's algorithm \cite{b9}. As illustrated in sub-figures b, c and d of Fig. \ref{severalpaths}, starting from the initial input edge, the algorithm traverses the hexagonal grid moving from one edge of a cell to all other edges in the cell, finding the least costly path. To dynamically account for updated costs due to changes in traffic, an ownship re-calculates its least costly path to its destination from its current position periodically. This enables it to choose the least costly path throughout the course of its trajectory. For all simulations conducted for this paper, an ownship recalculated its best path forward every time it entered a new hexagon.

\begin{figure}[hbt!]
\centering
\includegraphics[width=0.3\textwidth]{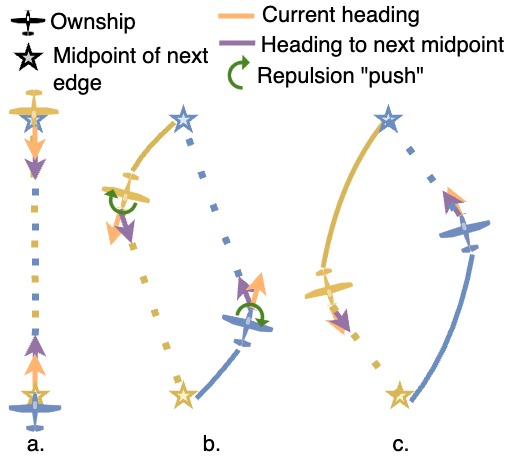}
\caption{(From left to right) Progression of a head-on scenario diversion where aircraft enter from opposite edges of a hexagon and head towards the edge the other aircraft entered from.}
\label{progress}
\end{figure}

\subsubsection*{Conflict resolution}
Since the ownship is navigating an airspace with multiple other vehicles, we utilize a non-communicative, negotiation-free algorithm for local collision avoidance \cite{b11}. 

In this setup, a repulsion modeled as an angular impulse ``pushes" the ownship away from other aircraft by making deviations to its heading. Consider two aircraft $m$ and $n$, in the North-East-Down coordinate frame. The instantaneous distance between them is given by ${r}_{mn}$ and the rate at which this distance between the two aircraft is changing is given by $\dot{r}_{mn}$. The instantaneous repulsion $R$ on an ownship $m$ is given by:

\begin{equation}
\label{gamma}
{R}_{m} = \sum_{n=1}^S  
\Bigg[ \frac{k_r}{|{r}_{mn}|^2} + 
k_{\dot{r}}\text{max}\Bigg(0, -\frac{\dot{r}_{mn} \cdot{r}_{mn}} {|{r}_{mn}|} \Bigg) \Bigg], \forall n \neq m
\end{equation} 
where $S$ is the total number of aircraft in the grid, and $k_r$ and $k_{\dot{r}}$ are the repulsion gain constants. An example of how this repulsion works is showcased in Fig. \ref{progress}. For this work, the conflict resolution for each aircraft is done sequentially and at every time step.

\begin{figure}[hbt!]
\centering
\includegraphics[width=0.46\textwidth]{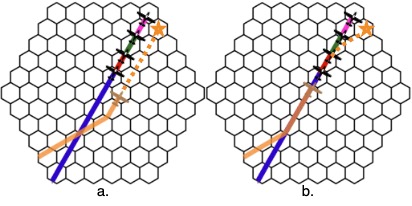}
\caption{Demonstrating the effect of $k_t$ for a single ownship: Aircraft are introduced into the airspace at staggered intervals, originating from the bottom-left with destinations in the top right. Only the last aircraft (orange) is designated to be an ownship here, i.e., has the option to follow or not. (a). $k_t$ = 0, i.e., no traffic-following is permitted. (b). $k_t$ = 3, i.e., traffic-following is permitted.}
\label{following_single_ownship}
\end{figure}

\subsubsection*{Updates to Ownship State}
Every time an ownship enters a hexagon, it determines the next best edge to travel to by using the path planning algorithm from its current position in the grid to its destination. Locally, it uses the collision avoidance algorithm to maintain horizontal separation from other aircraft in the system. Updates to an ownship's trajectory are made in the form of a cumulative heading change, $\Delta\psi$, (due to path planning and conflict resolution) while speed and altitude are kept constant. In a simulation with multiple ownships, each ownship's trajectory is updated sequentially, with no coordination or negotiation, at every time-step\cite{b11}. 

\subsection*{Examples} 
\label{subsec:example}

We now demonstrate traffic-following and non traffic-following behavior using the airspace setup, cost functions and navigation techniques described above in two example scenarios: single-ownship and multiple-ownships. In the single-ownship case, we introduce a convoy of four aircraft (different colors) into the grid as shown in Fig. \ref{following_single_ownship}. They enter from the bottom left corner, and their destination is the top right corner. A fifth aircraft is offset (orange) and is the ownship, i.e, can either choose to follow the convoy or not. Fig.\ref{following_single_ownship} a. depicts the case for when $k_t$ is set to 0, and Fig.\ref{following_single_ownship} b. depicts the case for when $k_t$ is set to 3. 
\begin{figure}[hbt!]
\centering
\includegraphics[width=0.45\textwidth]{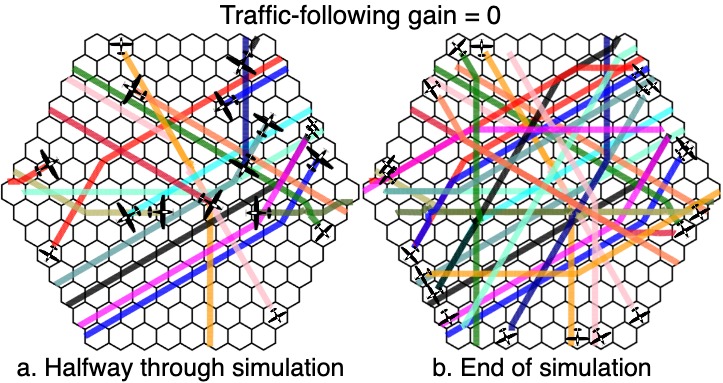}
\includegraphics[width=0.45\textwidth]{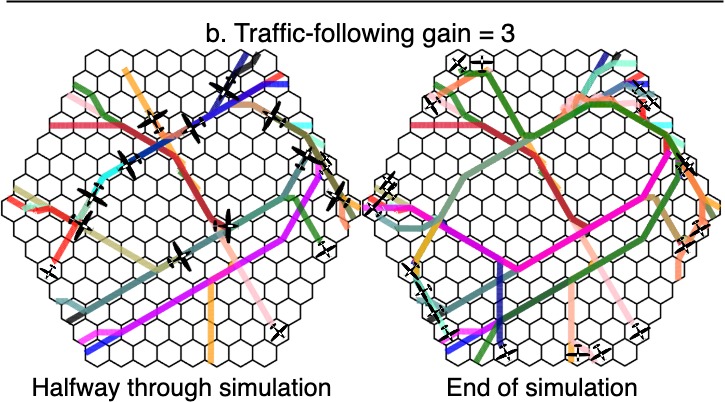}
\caption{
Demonstrating the effect of $k_t$ on traffic patterns when all aircraft are ownships: Order emerges when aircraft follow traffic patterns.}
\label{results1_1}
\end{figure}

Similarly, in the multi-ownship case depicted in Fig.\ref{results1_1}, all aircraft are ownships and have randomly assigned initial and final destinations on opposite sides of the grid. Fig. \ref{results1_1}a. shows the trajectories (at the  middle and end of the scenario) of aircraft for the case when $k_t = 0$, and Fig. \ref{results1_1}b. depicts when $k_t = 3$. Both of these examples show reduced disorder in the airspace when aircraft follow traffic patterns.

\subsection{Entropy}\label{entropy}
This section describes the utilization of entropy within the context of our paper. To measure the disorder in our system, we use a popular formula used to measure entropy in the field of information theory \cite{b10}. For a discrete random variable $\textit{X}$, that is distributed according to  ${\displaystyle p\colon {\mathcal {X}}\to [0,1]}$, 
the entropy is

\begin{equation}
    \label{shannon}
    {\displaystyle \mathrm {H} (X)=-\sum _{x\in {\mathcal {X}}}p(x)\log p(x)}
\end{equation}

where $x$ is a value from set $\textit{X}$, $p(x)$ is the probability of $x$ occurring in $\textit{X}$, and $\sum$ denotes the sum over the variable's possible values. 

It is important to note here that Eq.\ref{shannon} can be used for measuring the entropy of various factors in an airspace. Options include metrics such as the magnitude of total heading changes within an airspace, transit time for an aircraft, etc. For the scope of this analysis, we focus on studying differences in the number and directions of pathways that aircraft travel across an airspace.

More specifically, consider the airspace configuration composed of hexagonal cells, as discussed in previous sections. In the case of a single cell, we quantify entropy considering the following two factors: 1. the diversity of directions (entry-exit pairings) through which traffic flows and 2. the number of aircraft utilizing these pathways. As stated earlier, the traffic matrix, T, stores information about the number and directions of crossings between each entry-exit pairing in a cell over time. This information is then leveraged to calculate the entropy within that individual cell. Subsequently, by aggregating this uncertainty across all cells within a grid, we obtain a metric that encapsulates the overall entropy of the airspace.

Next, we discuss an example of measuring entropy in the context of the variety of directions in a cell. We evaluate entropy for the following cases, to see how well it corresponds with intuition. Consider three cases of 10 aircraft moving across a cell, one after the other. For the first case, all 10 aircraft traverse the same entry and exit pairing numbers, say 4 and 2. For the second case, 10 aircraft enter the airspace one after the other from the same edge (edge 4) but have two different options for their exit (edges 1 and 2). In the final case, assume all 10 aircraft enter and exit from a previously unused entry-exit pairing. 

Disorder, in terms of the number of directions, is most pronounced in the last case given the multitude of options for aircraft to choose entry-exit pairings. Conversely, it is lowest in the first case since each aircraft is restricted to a single edge pair. Now looking at this, for the first case, the traffic matrix, $T_1$, and the corresponding normalized matrix, $\hat{T}_1$, are: 

$$T_1 = \begin{bmatrix}
    0 & 0 & 0 & 0 & 0 & 0\\
    0 & 0 & 0 & 0 & 0 & 0\\
    0 & 0 & 0 & 0 & 0 & 0\\
    0 & 10 & 0 & 0 & 0 & 0\\
    0 & 0 & 0 & 0 & 0 & 0\\
    0 & 0 & 0 & 0 & 0 & 0\\
    \end{bmatrix};
\hat{T}_1 = \begin{bmatrix}
    0 & 0 & 0 & 0 & 0 & 0\\
    0 & 0 & 0 & 0 & 0 & 0\\
    0 & 0 & 0 & 0 & 0 & 0\\
    0 & 1 & 0 & 0 & 0 & 0\\
    0 & 0 & 0 & 0 & 0 & 0\\
    0 & 0 & 0 & 0 & 0 & 0\\
    \end{bmatrix}$$
The entropy for this cell, H(X), using Eq. \ref{shannon} is then  = $-(1\times \text{log(1)})$ = 0. The entropy in this case is 0 since it is deterministic, i.e., all aircraft only have one choice available. 
For the second case, the traffic matrix, $T_2$, and the corresponding normalized traffic matrix, $\hat{T}_2$, are:

$$
T_2 = \begin{bmatrix}
    0 & 0 & 0 & 0 & 0 & 0\\
    0 & 0 & 0 & 0 & 0 & 0\\
    0 & 0 & 0 & 0 & 0 & 0\\
    5 & 5 & 0 & 0 & 0 & 0\\
    0 & 0 & 0 & 0 & 0 & 0\\
    0 & 0 & 0 & 0 & 0 & 0\\
    \end{bmatrix}; 
\hat{T}_2 = \begin{bmatrix}
    0 & 0 & 0 & 0 & 0 & 0\\
    0 & 0 & 0 & 0 & 0 & 0\\
    0 & 0 & 0 & 0 & 0 & 0\\
    \frac{1}{2} & \frac{1}{2} & 0 & 0 & 0 & 0\\
    0 & 0 & 0 & 0 & 0 & 0\\
    0 & 0 & 0 & 0 & 0 & 0\\
    \end{bmatrix}
    $$
Therefore, the entropy, H(X), for this cell is $-(\frac{1}{2}\times \text{log}(\frac{1}{2}))\times(2)$ = 0.6931. For the third case, let's assume the traffic matrix, $T_3$, and the corresponding normalized traffic matrix, $\hat{T}_3$, are:

$$
T_3 = \begin{bmatrix}
    1 & 0 & 0 & 1 & 0 & 1\\
    0 & 0 & 0 & 1 & 0 & 0\\
    0 & 0 & 0 & 0 & 0 & 1\\
    1 & 1 & 0 & 0 & 0 & 0\\
    0 & 0 & 0 & 1 & 0 & 0\\
    0 & 1 & 0 & 0 & 1 & 0\\
    \end{bmatrix};
\hat{T}_3 = \begin{bmatrix}
    \frac{1}{10} & 0 & 0 & \frac{1}{10} & 0 & \frac{1}{10}\\
    0 & 0 & 0 & \frac{1}{10} & 0 & 0\\
    0 & 0 & 0 & 0 & 0 & \frac{1}{10}\\
    \frac{1}{10} & \frac{1}{10} & 0 & 0 & 0 & 0\\
    0 & 0 & 0 & \frac{1}{10} & 0 & 0\\
    0 & \frac{1}{10} & 0 & 0 & \frac{1}{10} & 0\\
    \end{bmatrix}
    \vspace{0.2cm}$$
H(X), for this cell is $-\frac{1}{10}\times \text{log}(\frac{1}{10})*10$ = 2.3026. Hence, the lowest entropy occurs in the first case, where aircraft all have the same entry-exit edges. The highest entropy is observed in the last case, where each aircraft has different entry-exit pairings. A similar approach can be used to understand entropy in the context of the number of aircraft traversing a cell. Between the following two traffic matrices case A has a higher entropy than case B.

$$T_A = \begin{bmatrix}
    0 & 0 & 0 & 0 & 0 & 0\\
    0 & 0 & 0 & 0 & 0 & 0\\
    0 & 0 & 0 & 0 & 0 & 0\\
    5 & 5 & 0 & 0 & 0 & 0\\
    0 & 0 & 0 & 0 & 0 & 0\\
    0 & 0 & 0 & 0 & 0 & 0\\
    \end{bmatrix};
 T_B = \begin{bmatrix}
    0 & 0 & 0 & 0 & 0 & 0\\
    0 & 0 & 0 & 0 & 0 & 0\\
    0 & 0 & 0 & 0 & 0 & 0\\
    1 & 9 & 0 & 0 & 0 & 0\\
    0 & 0 & 0 & 0 & 0 & 0\\
    0 & 0 & 0 & 0 & 0 & 0\\
    \end{bmatrix}
$$
This is because the uncertainty in the exit edge an aircraft will pick is higher in case A. 

\section{Computational Experiments}\label{sec:results}
In this section we present experiments that were conducted to demonstrate the emergence of order using the models and algorithms presented in the methodology. For these simulations, we introduce aircraft into a hexagonal grid using the same approach described in the example in Section \ref{subsec:example}. Each vehicle enters the airspace a certain time-period after the preceding one. This traffic-staggering approach allows for the accumulation of data in the traffic pattern map, which can be leveraged by subsequent aircraft for path planning. 

In this analysis, the size of each hexagonal cell is 2.65 miles and all aircraft are assumed to be traveling at a constant speed of 250 miles per hour. The repulsion constants, $k_r$ and $k_{\dot{r}}$, used in Eq. \ref{gamma}, are both set at 0.01, and the ``repulsion" only sets in when the horizontal distance between two aircraft is less than or equal to 10 miles. 

In the next two subsections, we examine the impact of traffic-following behavior and the amount of information available on airspace entropy. The third subsection delves into the effects of traffic-following behavior on aircraft travel times within an airspace and we finish the discussion section by discussing trade-offs for entropy gain versus travel times. 

\subsection{Effect of $k_t$ on airspace entropy} \label{sec:effects_k_t}

In this experiment, the objective is to study the effects of varying the traffic-following factor, $k_t$, on the entropy of an airspace. 

\begin{figure}[hbt!]
\centering
\includegraphics[width=0.4\textwidth]{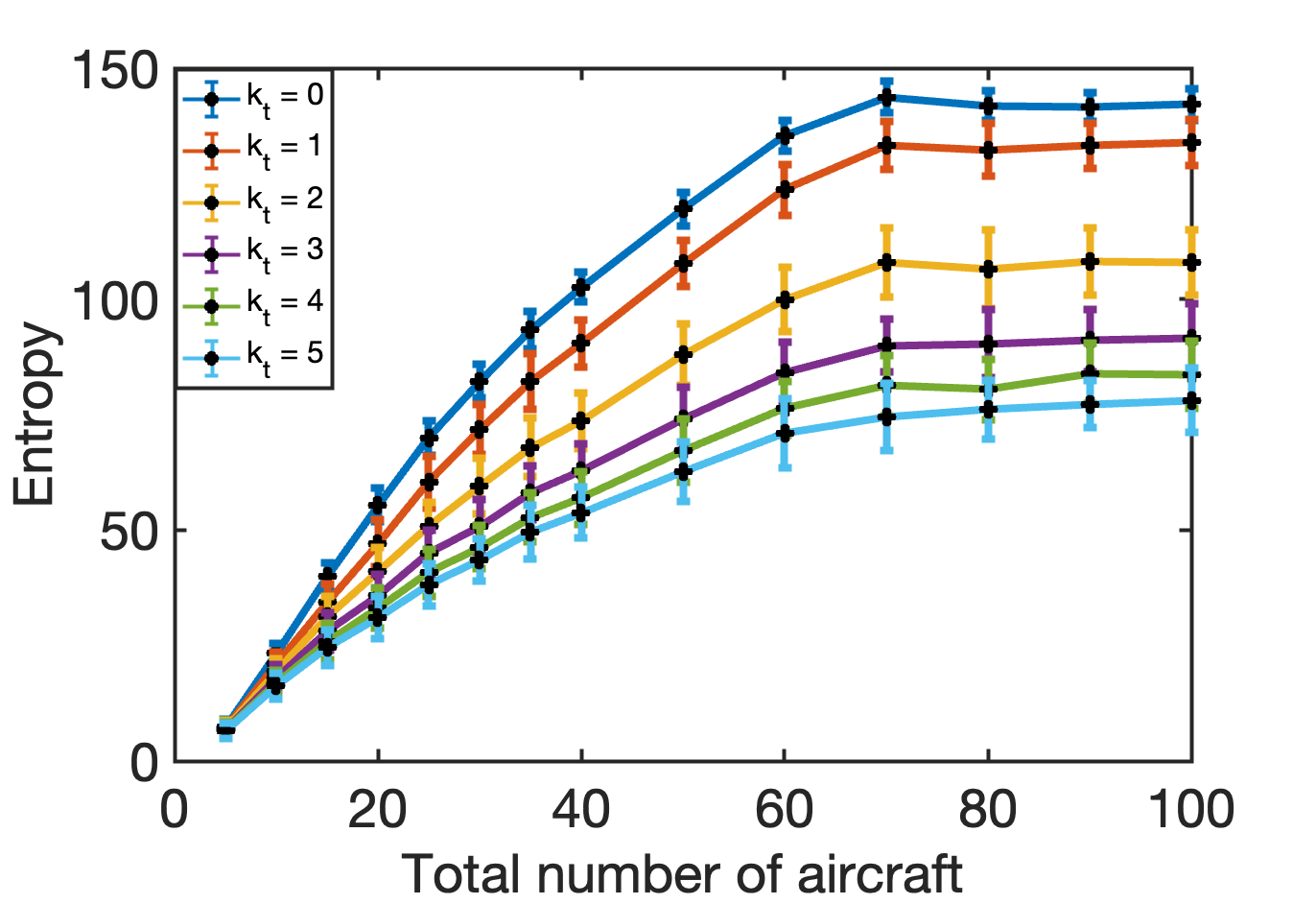}
\caption{Influence of the traffic-following factor, $k_t$, on airspace entropy: Analysis of 100 cases for each data point with staggered aircraft entries at fixed time intervals.}
\label{varying_k_t}
\end{figure}

For the hexagonal grid airspace configuration presented in Section \ref{sec:Method}, we explore various scenarios for each value of $k_t$. Specifically, we test values of $k_t$ on the same set of 100 scenarios, where a scenario is defined by randomly generated initial and final coordinates for different numbers of aircraft denoted as $N = \{5, 10, ... 100\}$. This ensures a fair comparison across different cases. In each scenario, aircraft enter the airspace sequentially, with a 40-second separation between each entry. This deliberate timing of 40 seconds is chosen because, in this setup, it takes an aircraft approximately 40 seconds to traverse a cell (based on the chosen cell size and aircraft speeds for this experiment) and begin contributing data to the traffic map. Subsequent aircraft entering the airspace can then leverage these insights into the flow of traffic within the airspace. This enables them to make informed decisions on the best path forward, depending on which cells are most cost-effective to traverse when taking into account prevailing traffic conditions and their assigned $k_t$. It should be noted here that, due to this staggered introduction of aircraft into the airspace, N is the total number of aircraft that have traversed the airspace over a given period of time, rather than the total number of vehicles at a specific instance in time. 

Results obtained from this experiment are shown in Fig.\ref{varying_k_t}. For the case of no traffic-following, i.e., when $k_t$ is set to 0, airspace entropy is highest. This agrees with intuition because when aircraft are not following other traffic, they are simply moving in the direction that suits them best, resulting in the traversal of an increasing number of new pathways in the airspace. For any given aircraft number, we see a decrease in the entropy of an airspace as the traffic-following factor increases. This corresponds to what one might expect intuitively, since as aircraft follow traffic, they are more likely to align their movements with other vehicles, naturally creating more pronounced ``airways" without the need for any predefined route structure. 

The anticipated leveling off of the curves at the end of the plots is due to the staggered entry method, explained at the beginning of Section \ref{sec:results}. As we move along the horizontal axis, the total number of aircraft input into the system is increasing, where some aircraft reach their destination by the time subsequent ones enter the airspace. Since subsequent aircraft entering simply follow pre-established pathways, they settle on a structure, and there is a resulting saturation in airspace entropy.

We observe that the marginal increase in traffic order (corresponding to a decrease in entropy) diminishes as we increase the traffic-following behavior. For example, a higher increase in airspace order is observed when $k_t$ is increased from 2 to 3 than when it is increased from 4 to 5. This is attributed to a higher number of aircraft following established pathways in the airspace when $k_t$ is increased from 2 to 3. In contrast, increasing the degree of traffic-following to 5 instead of 4 does not yield as substantial an increase in order, as the majority of aircraft are already following other traffic - once airways are established, all subsequent aircraft tend to adhere to the same pathways established by prior aircraft, with increasingly smaller added benefit in terms of more structure. 

It should be noted that this initial analysis is conducted in a deterministic setting and intended to demonstrate the emergence of traffic patterns and stable structure with traffic following. In future work we will add disturbances such as weather cells which would invoke a dynamic change in the emergent patterns. Under such uncertainties, the traffic pattern maps should be extended to be predictive over time and each agent traffic-following behavior would adapt dynamically to mitigate these uncertainties.

\subsection{Effect of $k_t$ on aircraft travel times}
In this experiment, we examine how aircraft travel time is affected while following other traffic. To analyze this, we monitor how long it takes each aircraft to get from its initial position to the destination using the same setup described in Section \ref{sec:effects_k_t}. 
\begin{figure}[hbt!]
\centering
\includegraphics[width=0.45\textwidth]{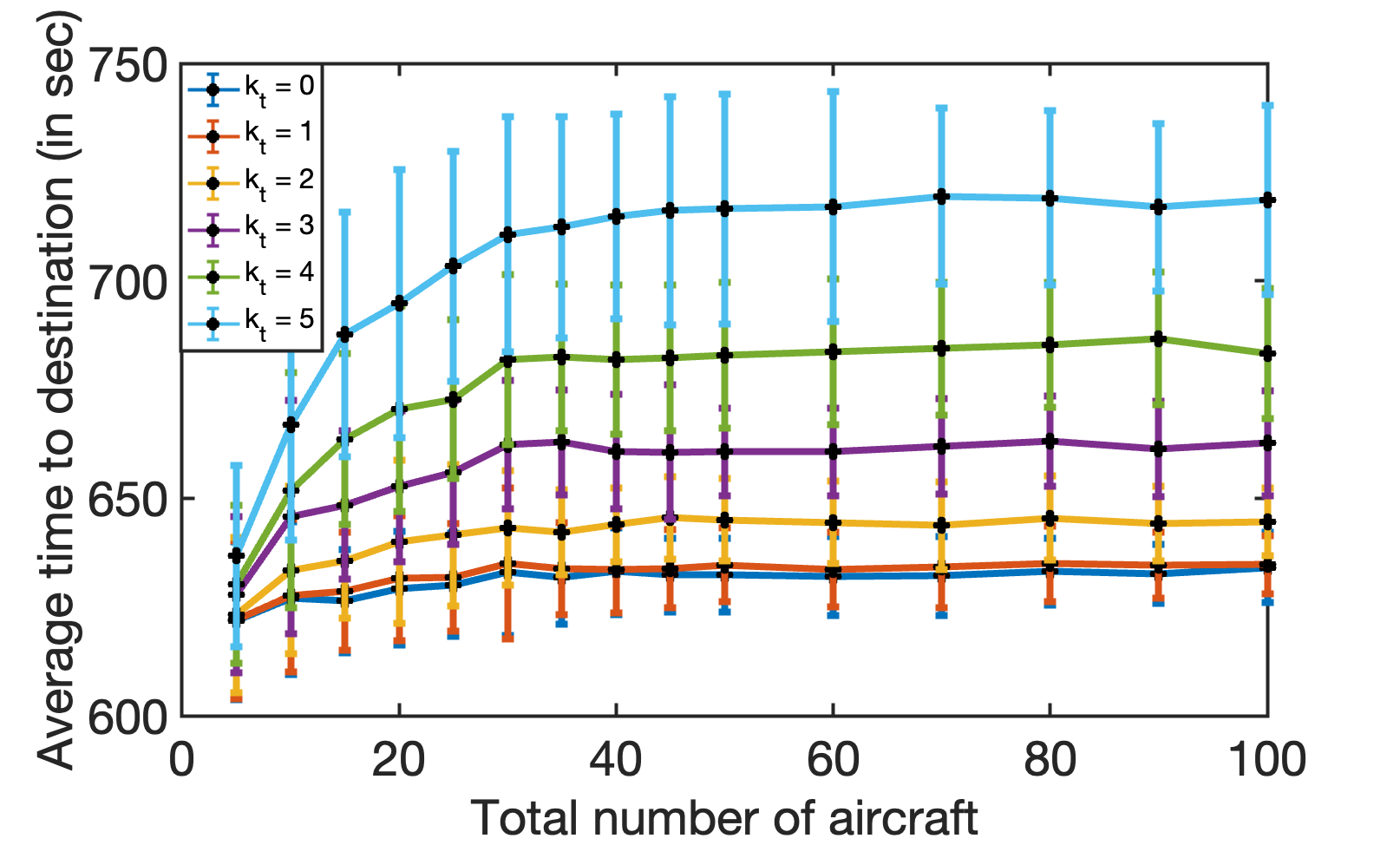}
\caption{The effect of $k_t$ on aircraft travel times.}
\label{time_to_destination}
\end{figure}

The results obtained from this experiment are shown in Fig.\ref{time_to_destination}. From this plot it is evident that, as expected, travel time increases when aircraft follow traffic patterns. As expected, increases in travel time are more pronounced with a higher $k_t$ as aircraft make more deviations to follow traffic. The plateau observed in these plots (starting at around 50 aircraft), can be attributed to the staggered introduction of aircraft into the system and the stabilizing of the traffic structure observed from the entropy saturation in Fig.\ref{varying_k_t}. We also observe an increase in the marginal effect of increasing traffic-following behavior on travel time; where increasing $k_t$ from 1 to 2 increases the travel time slightly, while increasing $k_t$ from 4 to 5 increases travel time much more.  

Note that in this experiment, some aircraft reach their destinations by the time subsequent ones enter the system. Given that the travel time remains almost at the same value in the case of no traffic-following, we can infer that we are observing aircraft travel times unaffected by congestion-induced increments. In future work, we intend to study the trade off between order and travel time under high density where congestion is in effect. 

\subsection{Trade-off between airspace entropy and aircraft travel times}

\begin{figure}[hbt!]
\centering
\includegraphics[width=0.4\textwidth]{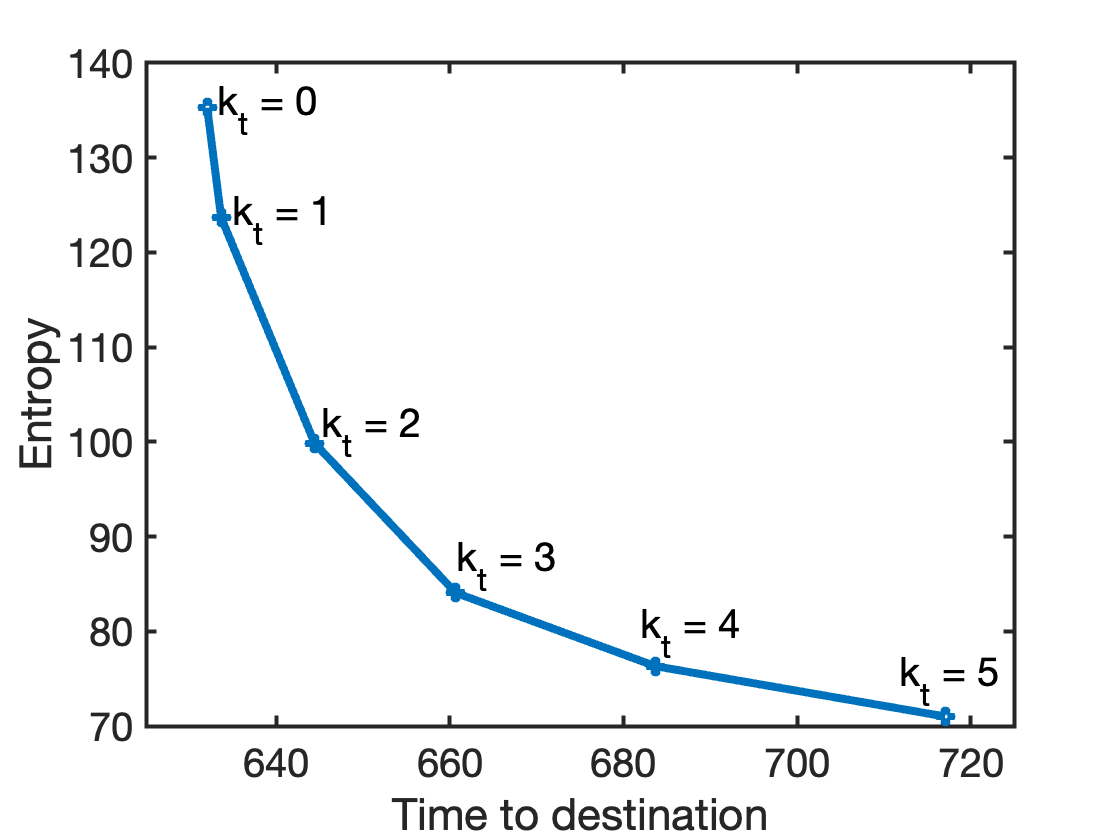}
\caption{Exploring the trade-off between entropy and travel time across different degrees of traffic-following behavior for 60 aircraft.}
\label{time_vs_entropy}
\end{figure}

In this section we combine the effects on entropy and travel time from Figs. \ref{varying_k_t} and \ref{time_to_destination}, respectively, under conditions where the traffic structure in the airspace has stabilized. Considering this occurs at approximately 60 aircraft, as discussed in the previous subsection, we select this number of aircraft for the plot shown in Fig.\ref{time_vs_entropy}. For scenario with a total number of aircraft exceeding 60, similar plots are observed, as both the entropy and travel time values exhibit a plateau. The trade off between entropy and travel time is clear in the figure, where it is observed that for low values of the traffic-following constant, such as $k_t$ = 1, 2, the increases in travel time are minimal with notable gains in entropy. The reverse is observed at high $k_t$ values where additional traffic following behavior results in excessive penalty in terms of travel time with minimal additional order benefits. The plot demonstrates how the methods presented in this paper can be used to select appropriate traffic-following gains to trade off order against other priorities. 

\section{Conclusions and Future Work}\label{sec:future}

In this paper, we investigated the emergence of traffic order in a distributed autonomous multi-vehicle system and its effect on travel time. To conduct this study, we constructed a dynamically updating traffic pattern map of the airspace based on the directions of traffic and the number of aircraft in these traffic flows. We developed an algorithm for each aircraft to adjust the degree to which it follows traffic patterns in the map versus other priorities such as travel time. Our analysis of entropy effects for varying levels of traffic-following revealed that, for low traffic-following behavior, there is minimal penalty for average travel times, accompanied by simultaneous benefits in improving airspace orderliness. Conversely, continuing to increase traffic-following behavior results in prolonged travel times, yet produces marginal reduction in the overall entropy of the airspace.

For the next chapter of this work, insights gained at low densities considered in this work are being extrapolated to investigate the trade off between order and travel time under high density situations. By developing a methodology to study the importance of order in distributed, multi-agent systems, we have taken an initial step that can be extended to broader metrics for collective autonomous aerial systems. While this preliminary work was deterministic, future extensions will explore the dynamic emergence of order under more-realistic dynamic uncertainties such as weather disturbances and limited agent information about the traffic. This entails incorporating predictive techniques for the traffic pattern estimation and robust agent behavior to mitigate the effects of uncertainty. Overall, this methodology can be used to determine the ideal traffic-following factor for each agent, taking into consideration the relative importance of travel times and airspace entropy. This will ultimately create airspace order with as few rules and as little structure as possible.

\end{document}